\documentclass[twocolumn,prb,superscriptaddress,aps,showpacs]{revtex4-1}
\usepackage{graphicx}
\usepackage{amsmath}
\usepackage{multirow}
\usepackage{physics}
\usepackage{color}
\usepackage{bm}
\newlength{\figwidth}
\newlength{\figwidthb}
\newcommand{\bk}{\textbf{k}}
\newcommand{\nn}{\langle ij \rangle}
\newcommand{\nnn}{\langle\langle ij \rangle\rangle}
\newcommand{\omn}{\omega_n}

\newcommand{\be}{\begin{equation}}
\newcommand{\ee}{\end{equation}}

\newcommand{\bp}{{\bf p}}

\newcommand{\bq}{{\bf{q}}}

\newcommand{\bea}{\begin{eqnarray}}
\newcommand{\eea}{\end{eqnarray}}
\newcommand{\beal}{\begin{align}}
\newcommand{\eeal}{\end{align}}

\newcommand{\ra}{\rangle}
\newcommand{\la}{\langle}

\newcommand{\dg}{{\dagger}}
\newcommand{\pdg}{{\phantom\dagger}}

\begin{document}

\title{Bilayer Haldane model: From trivial band insulator to fractionalized quantum anomalous Hall insulator}

\author{Sopheak Sorn}
\affiliation{Department of Physics, University of Toronto, Toronto, Ontario M5S 1A7, Canada}
\date{\today}
\begin{abstract}
Motivated by work on the bulk topological proximity effect and the topological bootstrap, we consider two coupled layers of quantum anomalous Hall (QAH) insulators with
opposite signs of time-reversal breaking, which leads to a trivial band insulator at half-filling. We study the impact of interactions in this model within slave rotor
theory, which leads to a layer-selective Mott transition, resulting in a fractionalized quantum anomalous Hall insulator QAH$^*$ where a Chern band insulator coexists with a chiral spin liquid.
We also compute the edge electron spectral function in the vicinity of the QAH$^*$ phase.
\end{abstract}

\maketitle

\section{\label{sec:introduction}Introduction}
Recent theoretical work on topological phases of matter has introduced the concept of a ``bulk topological proximity effect'' (BTPE)\cite{TPE}, wherein a topologically trivial layer coupled to 
topologically nontrivial bands \cite{TI_rev} 
can itself exhibit nontrivial topological character of the {\it opposite} type. This arises from virtual hopping transitions into the nontrivial layer.
An interesting variant of this idea, which was subsequently explored, is the ``topological bootstrap''\cite{bootstrap}, where isolated spins can be driven into a topologically ordered chiral spin liquid
phase \cite{CSL1, CSL2, topo_order, CSL3, CSL4} via Kondo coupling to nontrivial Chern bands\cite{QAHE}. 
Naively, we expect that increasing the strength of the Kondo coupling might lead to a transition into a trivial insulator, where every spin binds an
electron.

In a different research trend studying the effects of electron correlation on band topology, new correlation-driven phases of matter have been found with and without spontaneously broken symmetries\cite{review_fractionalized_ti1, review_fractionalized_ti2, review_tki}, e.g. antiferromagnetic Chern insulators\cite{xidai} and fractionalized topological insulators with neutral gapless surface excitations\cite{pesin, rachel}.
Motivated by the rich physics of such phases and
to explore the connection between BTPE and the topological bootstrap, we study a toy bilayer Haldane model, where each layer hosts spin-$1/2$ electrons in topologically
nontrivial phases but which are of the opposite type. At half-filling, the total Chern number of the occupied ``valence'' bands is then zero. We assume that one of the layers could
have a bandwidth reduced by a factor $0 < \lambda < 1$. For small $\lambda \ll 1$, we may view one of these layers as having inherited its 
nontrivial ``opposite'' band topology due to the BTPE. In this setting, we study how tuning Hubbard 
interactions in one layer eventually leads to a chiral spin liquid Mott insulator which effectively decouples from the other layer, so that the net combination acts as a fractionalized quantum anomalous Hall insulator QAH$^*$, which has bulk semions and topological order coexisting with a quantum Hall effect. 
Such a system would have chiral charge edge modes and a counterpropagating neutral edge mode, so that it would exhibit a quantized thermal Hall effect and a quantized electrical Hall effect which violate the Wiedemann-Franz law.
For $\lambda \ll 1$, this QAH$^*$ phase is identifiable as that obtained within the topological bootstrap picture.

QAH$^*$ is analogous to the fractionalized Fermi liquid FL$^*$ studied in Ref.~\onlinecite{frac_FL}, in which a spin liquid coexists with a Fermi liquid. It is distinct from a correlated Chern insulator phase CI$^*$\cite{ci_starred} obtained within the slave spin theory of a single-layer Chern insulator. The CI$^*$ phase also possesses fractionalized quasiparticles and exhibits a quantized charge Hall effect, yet it has no electron-like quasiparticles in contrast to QAH$^*$. In the language of Ref.\onlinecite{orthogonal_fl}, the CI$^*$ may be viewed as an ``orthogonal" QAH insulator.

We have 
explored the phase diagram of this bilayer model within a slave rotor mean field theory calculation\cite{Florens2002, Florens2004, SR01, SR02, SR03}. In the presence of inversion symmetry, the ``valence'' bands
have Chern numbers $\pm 1$, while incorporating inversion breaking terms renders each ``valence'' band to individually have Chern number zero. 
As opposed to the Kondo lattice model explored in the context of the 
topological bootstrap, this model can be viewed as a periodic Anderson model in which a layer-selective Mott transition
leads to the QAH$^*$ phase. 
However, simply increasing correlations on one layer does not necessarily directly drive the system into the QAH$^*$ phase since the correlated 
bands tend to drift up in energy with increasing interaction strength and thus get progressively depopulated; we thus generically
need an additional bias potential in order to convert the correlated layer into a half-filled Mott insulator.
Finally, in addition to the above discussed trivial band insulator and QAH$^*$ phases, we
find wide regimes of Chern metal and new Dirac semimetal phases.

We compute the edge electron spectral function in the correlated trivial band insulator as we approach the QAH$^*$ phase. Deep in the trivial
band insulator regime, there are counterpropagating electronic edge modes which hybridize and gap out. However, closer to the QAH$^*$ phase, this
hybridization strongly decreases. Furthermore, we find that 
while the chiral edge mode emanating from the noninteracting layer has high spectral intensity, the counterpropagating edge mode has a diminished intensity
which is progressively weakened upon approaching the layer Mott transition. Thus, although the correlated band insulator is topologically
trivial, its spectral function signature close to the QAH$^*$ phase
may be (incorrectly) suggestive of a topologically nontrivial state. In the QAH$^*$ phase, the edge modes decouple.

 The phase diagram is summarized in Fig.~\ref{fig:pd} and is the focus of the rest of this paper. Before we turn to this, it is important to note
 that while our results indicate the type of phases which might arise in the presence of correlations,
 additional interactions may be needed to stabilize the QAH$^*$ phase in this specific microscopic model\cite{CSL_Haldane_Hubbard} once 
 we allow for competition with spontaneous magnetically ordered phases. The paper is organized as follows. Section \ref{sec:model} outlines the
 model Hamiltonian and its symmetries. Section \ref{sec:nonint} discusses its non-interacting phase diagram. Section \ref{sec:SR} discusses the results from a slave rotor
 theory of the interactions, using a non-linear sigma model approach to the rotor fluctuations, and presents numerical results for the edge electron spectral function.
 
\section{\label{sec:model}Model and Symmetries}

\subsection{Bilayer Hamiltonian}

\begin{figure}[t]
  \includegraphics[width=0.48\textwidth]{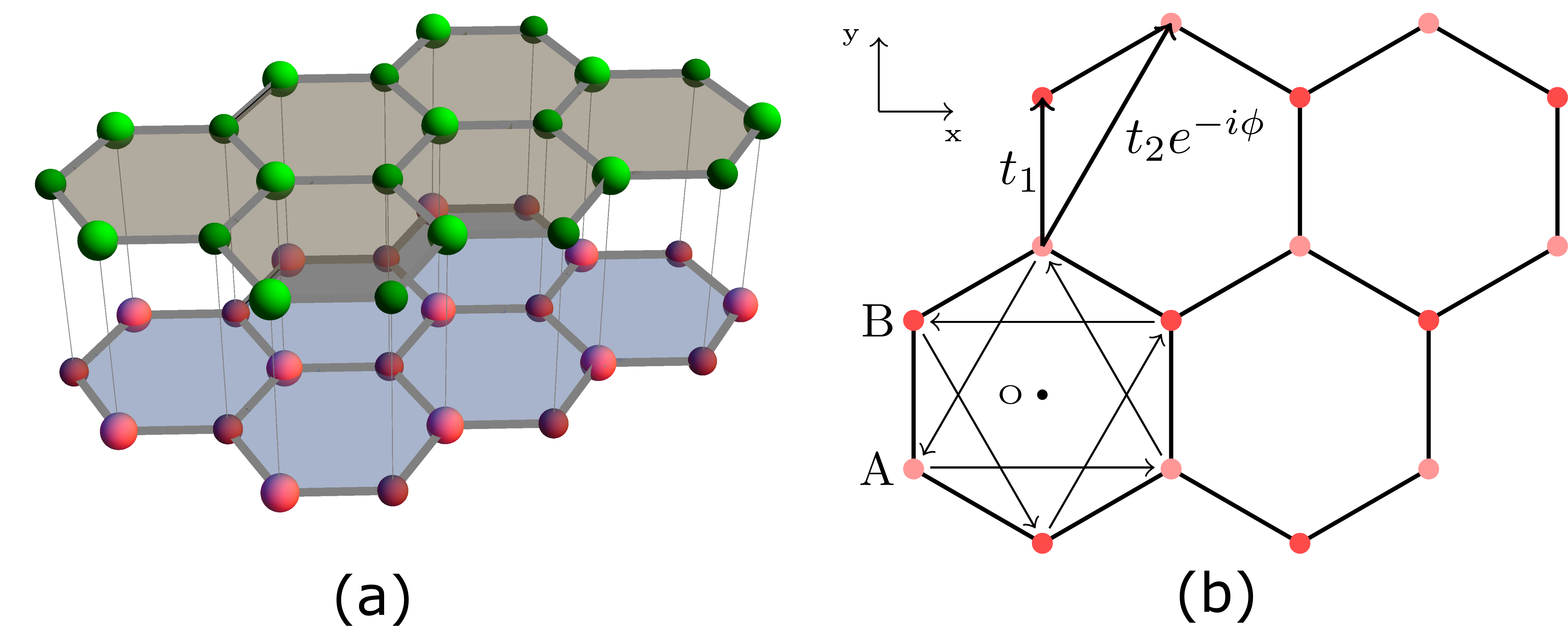}
\caption{(a) Bilayer honeycomb lattice showing vertical AA stacking. (b) Bottom layer (layer-$2$) depicting first and second neighbor hoppings $t_1$ and $t_2$. 
The second neighbor hopping is complex, given by  $t_2 e^{-i \nu_{ij} \phi}$, and arrows in the lowest left plaquette denote the directions of 
positive $\nu_{ij} = +1$, where $\nu_{ji}=-\nu_{ij}$.  The top layer (layer-$1$) has reversed sign of $\nu_{ij}$.}
\label{fig:bilayer}
\end{figure}

The bilayer Hamiltonian we study consists of three parts: a Haldane model on each layer, an interlayer hopping term which hybridizes the Chern bands
of the individual layers, and a Hubbard interaction which drives a Mott transition on one layer, so that
\bea
\label{eq:model-hamiltonian}
H &=& H_1^\pdg + \lambda H_2^\pdg + H_{\rm hyb} +  H_2^{\rm int}.
\eea	
Here, $H_1$ and $H_2$ denote the noninteracting hopping Hamiltonian on layers $1,2$ respectively. The bands of layer-$2$ are rescaled with respect to layer-$1$ by a parameter $\lambda \!\!<\!\! 1$. Let us 
denote the electron operators on layer-$1$ and layer-$2$ as $c^{\dagger}$ and $d^{\dagger}$ respectively.
The noninteracting layer-$1$ Hamiltonian $H_1 \equiv H_{\rm Haldane}(t_1, t_2, \phi, M)$ is the honeycomb lattice 
Haldane model Hamiltonian\cite{Haldane_model}
\bea
H_{\rm Haldane} &=& 
\!-\! \sum_{\nn, \sigma} t^\pdg_1 c^{\dagger}_{i\sigma}c^\pdg_{j\sigma} \!-\! \sum_{\nnn, \sigma} t^\pdg_2 e^{-i\nu_{ij}\phi} c^{\dagger}_{i\sigma}c^\pdg_{j\sigma} \!+\! \text{H.c.}
\nonumber \\
&+& M\sum_{i\sigma}\epsilon^\pdg_i c^{\dagger}_{i\sigma}c^\pdg_{i\sigma},
\label{eq:haldane}
\eea
while $H_2 \equiv H_{\rm Haldane}(t_1, t_2, -\phi, M)$ with fermion operators $c^\dg \to d^\dg$.
Here, $\sigma = \uparrow, \downarrow$ labels spin, and $\epsilon^\pdg_i = \pm 1$ labels the respective sublattices $A$ and $B$ so that $M$ controls the breaking of the 2D inversion 
symmetry.
The second-neighbor hopping term breaks time reversal symmetry (TR).
As shown in Fig.~\ref{fig:bilayer}, $\nu_{ij} = \pm 1$, which results in an alternating flux profile with a vanishing total flux through each hexagon. Here and
below, we will set $t_1=1$.

The hybridization Hamiltonian $H_{\rm hyb}$ encapsulates interlayer hopping (which is momentum-independent), and
a layer bias potential $\Delta$:
\bea
  \label{eq:hybridization}
    H_{\text{hyb}} &=& - t_{\perp} \sum_{i\sigma} \left( c^{\dagger}_{i
  \sigma}d^\pdg_{i\sigma} + \text{H.c.} \right) - \Delta \sum_{i\sigma} d^{\dagger}_{i\sigma}d^\pdg_{i\sigma}.
\eea
Electron-electron interactions are encoded in $H_2^{\rm int}$, which is the on-site Hubbard repulsion; for simplicity, we have assumed that this
interaction is only present on layer-$2$:
\bea
  \label{eq:haldane-hubbard}
    H_2^{\rm int} &=& U\sum_i n_{d, i, \uparrow}n_{d, i,\downarrow},
\eea
  where $n_{d, i, \sigma} = d^{\dagger}_{i\sigma}d_{i\sigma}$.  Such a Hubbard repulsion will drive a Mott transition in layer-$2$. When $\lambda < 1$, meaning 
 the two layers are {\it inequivalent}, with layer-$2$ having a smaller bandwidth, turning on a Hubbard interactions in both layers will drive a similar
 layer-selective Mott transition in layer-$2$.
  
\subsection{Symmetries}
The bilayer Haldane model has the following symmetries: (1) translational symmetry of the honeycomb lattice;
(2) $C_3$ spatial rotation symmetry about the center of each hexagonal plaquette; (3) $SU(2)$ spin rotation symmetry; 
(4) while time reversal symmetry ${\cal T}$ (i.e., complex conjugation which reverses flux $\phi \to -\phi$) is broken, ${\cal T M}$ 
which combines it with a mirror operation ${\cal M}$ is a good symmetry.
Here, the mirror line connects opposite vertices of the hexagon.
(5) Finally, when $M=0$, there is 2D inversion symmetry, which is
equivalent to $\pi$ rotation about the hexagon center; it 
sends the momentum $\bk \mapsto -\bk$ and exchanges the two sublattices.

\section{\label{sec:nonint} Noninteracting phase diagram}

Before studying the effects of interaction, we compute the noninteracting phase diagram of the bilayer.

When $t_\perp=0$, the two layers are decoupled. In this limit,
for $t_2, \phi \neq 0$ and when $|M/t_2| < 3\sqrt{3} |\sin \phi|$, each band in each layer carries a nontrivial Chern number, resulting in a 
quantum anomalous Hall effect at half filling, while larger $|M/t_2|$ results in a trivial band insulator\cite{Haldane_model}.
However, the fact that the phase $\phi$ in layer-$2$ is negative of that in layer-$1$ renders the whole system topologically 
trivial even when $M=0$.

Here, and below, we fix the density to be at half filling, and set $t_2 = 0.25, t_{\perp} = 0.3$ and $\phi = \pi/2$, and
explore the phase diagram as we tune $M$ and $\Delta$ for $\lambda=0.5$. We discover three phases.

(1) A trivial band insulator such that
the total Chern number of occupied bands is zero. However,
bands below the Fermi level (i.e., valence bands) may carry individually either nonzero Chern numbers or zero Chern number.
We notice that individually trivial valence bands with Chern numbers $(0, 0)$ can only be achieved when $M \neq 0$.

 (2) As one increases the strength of $\Delta$, the band insulator gives way to a Dirac semimetal with $6$ band touching points in the Brillouin zone (BZ).
 The Dirac cones emerge pairwise from each M point, and move towards towards the $\Gamma$ point. 
 When $M = 0$, the Dirac cones are situated perfectly on the $\Gamma$-M lines but are otherwise rotated away. 
 This can be understood as a mathematical structure of the Bloch Hamiltonian; see Appendix \ref{appdx: origin-dirac-points}. 
 The Dirac cones are, however, not protected by any symmetry and can be gapped out, for instance, by introducing a third-nearest neighbor hopping.
 
(3) Upon increasing $M$, the phase diagram shows an ambipolar metallic phase with no Dirac band touching points, 
but with electron pockets and hole pockets around the K and K' points respectively.

\begin{figure}[t]
  \includegraphics[width=0.425\textwidth]{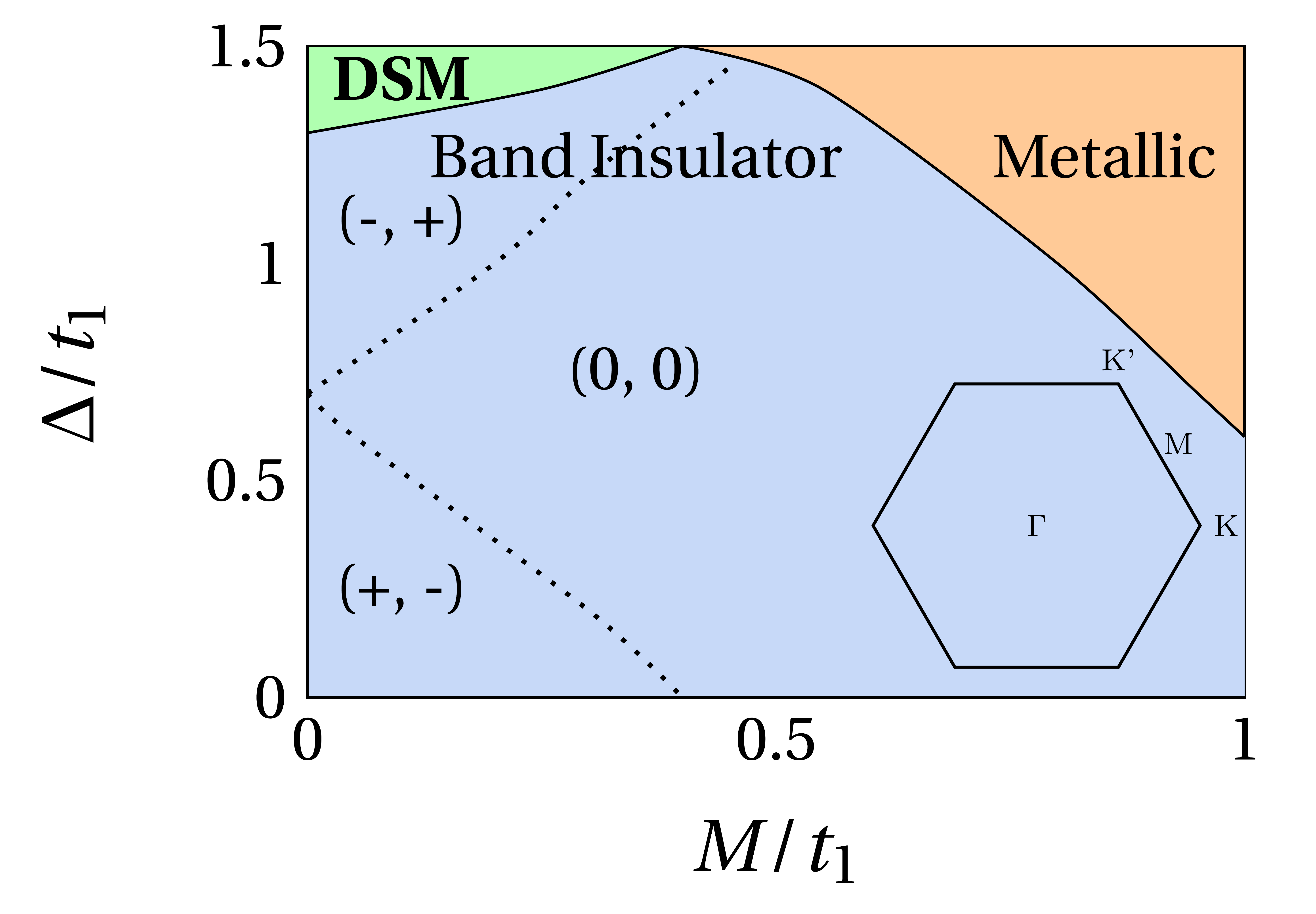}
\caption{Noninteracting phase diagram for $t_{\perp} = 0.3t_1$ and $\lambda = 0.5$. Band insulator is subdivided into three categories characterized by the Chern numbers of the valence bands. The $(0, 0)$ phase dominates and is present only with a nonzero $M.$ The Dirac semimetal has 6 Dirac cones in the BZ with the Fermi level at the Dirac points. These Dirac
points can be gapped out by further hopping, e.g. third-nearest neighbor hopping. 
The metallic phase arises from the effect of large $M$ which causes the band to have extrema in the vicinity of the K and K' points. 
At half filling, the band structure has electron pockets around K points and hole pockets around K' points.}
\label{fig:noninteracting}
\end{figure}

In the rest of the paper, we use slave rotor theory to
study the effect of interactions on the phase diagram in Fig.~\ref{fig:noninteracting}. We will
mainly explore the impact of varying $U$ and $\Delta$, for different values of $M$, starting from noninteracting phases
which are predominantly topologically trivial. This corresponds to starting from vertical cuts through the noninteracting phase diagram
and varying $U$.

\section{\label{sec:SR}Slave rotor mean field theory}
\subsection{Slave rotor representation}
Slave rotor representation has been used in studying Mott insulating phases and Mott 
transitions in strongly correlated systems.\cite{Florens2002, Florens2004, SR01, SR02, SR03}
Here we make use of this representation to study a layer-selective 
Mott transition in the interacting bilayer model.
In this representation, the electron operator in the correlated layer-$2$ is decomposed as 
$d^{\dagger}_{i\sigma} = f^{\dagger}_{i\sigma}e^{-i\theta_i}$,
into a fermionic spinon operator $f^\dg_{i\sigma}$ and a rotor operator $e^{-i\theta_i}$ which 
respectively carry the spin and charge degrees of freedom of the electron.
To project this expanded Hilbert space back to the physical electron Hilbert space, we need
to impose the local constraint
\be	
\label{eq:hilbert-constraint}
n_{f i} + L_i - 1 = 0.
\ee
Electron hopping terms in $H_2$ can be recast in the form $f^{\dagger}_{i\sigma}f_{j\sigma}e^{-i\theta_i}e^{i\theta_j}$, while 
the hybridization term becomes
\bea
H_{\text{hyb}} &=&  -\sum_{i\sigma} t^\pdg_{\perp}f^{\dagger}_{i\sigma}e^{-i\theta_i}c^\pdg_{i\sigma} + \text{H.c.} - \Delta \sum_{i\sigma} f^{\dagger}_{i\sigma}f_{i\sigma}.
\eea
The Hubbard interaction term is written as
\bea
H_2^{\rm int} = \frac{U}{2} \sum_{i} \left(L_i^2 + n_{f i} - 1\right),
\eea
where we have used the relation
\be
n_{d  i  \uparrow}n_{d  i \downarrow} = n_{d i}(n_{d i} - 1)/2 = (L_i^2 + n_{f i} - 1)/2,
\ee 
which is valid when the constraint in Eq.~(\ref{eq:hilbert-constraint}) is obeyed.

\subsection{Mean field theory}
	To make progress, we consider the following mean field decoupling of the spinon-rotor interaction terms.
\bea
	  f^{\dagger}_{i\sigma}f_{j\sigma}e^{-i\theta_i}e^{i\theta_j} &\approx& \langle f^{\dagger}_{i\sigma}f_{j\sigma}\rangle e^{-i\theta_i}e^{i\theta_j} + f^{\dagger}_{i\sigma}f_{j\sigma} \langle e^{-i\theta_i}e^{i\theta_j}\rangle \nonumber \\
	  &-& \langle f^{\dagger}_{i\sigma}f_{j\sigma} \rangle\langle e^{-i\theta_i}e^{i\theta_j}\rangle\\
	  f^{\dagger}_{i\sigma}c_{i\sigma}e^{-i\theta_i} &\approx& \langle f^{\dagger}_{i\sigma}c_{i\sigma} \rangle e^{-i\theta_i} + f^{\dagger}_{i\sigma}c_{i\sigma} \langle e^{-i\theta_i}\rangle \nonumber \\
	  &-&  \langle f^{\dagger}_{i\sigma}c_{i\sigma} \rangle \langle e^{-i\theta_i} \rangle,
\eea
where the expectation values $\langle \cdots \rangle$, dubbed ``bond mean fields'', are to be determined self-consistently. This decoupling scheme splits the 
Hamiltonian into two parts: one involving coupled spinons and $c$-electrons, and the other involving rotors. Equivalently, the
many-body electron wavefunction is then of the form
\begin{align}
\ket{\Psi_{\rm MF}} &= \ket{\Psi_{fc}} \otimes \ket{\Psi_{\theta}},
\end{align}
where $\ket{\Psi_{fc}}$ is the coupled spinon and $c$-electron wavefunction and $\ket{\Psi_{\theta}}$ is the rotor wavefunction,
with the constraint in Eq.~(\ref{eq:hilbert-constraint}) being imposed on average.

Here, we will focus on mean field ground states which do not break any symmetries of the model Hamiltonian, 
so we consider a ``uniform" ansatz. In this case, the bond mean fields are parametrized by only a few parameters. For nearest-neighbor bonds,
\bea
\label{eq:uniform-ansatze-nn}
\langle f^{\dagger}_{i\sigma}f_{j\sigma} \rangle &=& F_{nn}\\
	    \langle e^{-i\theta_i}e^{i\theta_j} \rangle &=& X_{nn},
\eea
	where $F_{nn}$ and $X_{nn}$ are real-valued and identical on all bonds 
	due to the combination of translation, $C_3$, and ${\cal T}{\cal M}$ symmetries. (Note that
	in the slave 
	rotor representation, ${\cal T}$ sends $f_{i\sigma} \to f_{i\sigma}$, $e^{\pm i \theta_i} \to e^{\pm i \theta_i}$,
	and conjugates complex numbers).
For next-nearest neighbors, there are two distinct bond mean fields corresponding to the two sublattices:
\bea
\langle f^{\dagger}_{i\sigma}f^\pdg_{j\sigma} \rangle &=& F_{nnn, A(B)}e^{-i\nu_{ij}\varphi_{A(B)}}\\
\langle e^{-i\theta_i}e^{i\theta_j} \rangle &=& X_{nnn, A(B)}e^{-i\nu_{ij}\eta_{A(B)}}.
	\label{eq:uniform-ansatze-nnn}
\eea
The bond mean fields for the interlayer term are
\bea
\langle f^{\dagger}_{i\sigma}c^\pdg_{i\sigma}\rangle &=& F_{\perp, A(B)}\\
\langle e^{i\theta_i} \rangle &=& \bar{X}_{A(B)}.
\label{eq:uniform-ansatze-hyb}
\eea
They can be chosen to be real-valued.
We impose the constraint (\ref{eq:hilbert-constraint}) on average by introducing two Lagrange multipliers, $\lambda_A$ and $\lambda_B$, 
for the two sublattices. The mean field theory now amounts to self-consistently
solving separate rotor and coupled spinon-$c$ Hamiltonians, $H_\theta$ and $H_{fc}$ respectively.
	
\subsection{Fermionic and rotor Hamiltonians}

	The fermionic part of the mean field Hamiltonian, involving $c$ and $f$, is given by
\bea
	    H_{fc} &=& H_1 + \lambda H_{2, f} + H_{\text{hyb}, f} + \frac{U}{2} \sum_{i} n_{f, i} \nonumber \\
	    &+& \sum_{i} \lambda_i n_{f, i} -\mu \sum_{i} (n_{f, i} + n_{c, i}),
\eea
	where the second last term comes from the constraints (\ref{eq:hilbert-constraint}), and the last term is the chemical potential, used to impose the electron density at half filling. $H_1$ is unaltered, while	the rest is given below.
\bea
	  \label{eq:def-terms-in-hf}
        H_{2, f} &=& - \sum_{\nn, \sigma} t_1X_{nn}f^{\dagger}_{i\sigma}f_{j\sigma} + \text{H.c.} \nonumber \\
        &-& \sum_{\nnn \sigma} t_2 X_{nnn, i} e^{-i\nu_{ij}\eta_i}e^{-i\nu_{ij}\phi}f^{\dagger}_{i\sigma}f_{j\sigma} + \text{H.c.} \nonumber \\
        &+& M\sum_{i\sigma}\epsilon_i f^{\dagger}_{i\sigma}f_{i\sigma}\\
        H_{\text{hyb}, f} &=& - \sum_{i\sigma} \left( t_{\perp} \bar{X}_i f^{\dagger}_{i\sigma}c_{i\sigma} + \text{H.c.}\right)  - \Delta \sum_{i} n_{f, i}.
\eea
We can now compute the ground state and the expectation values in (\ref{eq:uniform-ansatze-nn})-(\ref{eq:uniform-ansatze-hyb}) in order to solve 
the fermionic sector, and then evaluating averages $F_{nn}$ and $F_{nnn}$.

	The rotor Hamiltonian is given by
\bea
	\label{eq:rotor-mf-hamiltonian}
	    H_{\theta} &=& - \sum_{\nn} 2\lambda t_1F_{nn} X^{\dagger}_iX_{j} + \text{H.c.} \nonumber \\
	    &-& \sum_{\nnn} 2\lambda t_2 F_{nnn,i} e^{-i\nu_{ij}(\phi + \varphi_i)}X^{\dagger}_iX_{j} + \text{H.c.} \nonumber \\
	    &-& \sum_i 2t_{\perp} F_{\perp,i} X^{\dagger}_i \!+\! \text{H.c.} \!+\! \frac{U}{2} \sum_i L_i^2 \!+\! \sum_i \lambda_i L_i,
\eea
	where the operator $X_i \equiv e^{i\theta_i}$. The factors of 2 arise from spin sums in the spinon sector.
	
	To solve for the ground state expectation values in (\ref{eq:uniform-ansatze-nn})-(\ref{eq:uniform-ansatze-hyb}), we 
	integrate out the angular momentum and resort to a nonlinear sigma model representation of the rotor Hamiltonian which we treat 
	at Gaussian level as an approximation. 
	This last step involves solving a quadratic action in the sigma field, which can then be used to compute the bond mean fields 
	(see Appendix \ref{appdx: sigma-model} for more details).

A useful quantity in this approach is the expectation value $\langle X_i \rangle$ which distinguishes a Mott insulating phase from a non-Mott phase. 
When $\langle X_i \rangle$ vanishes, the charge fluctuation is strongly suppressed, which entails a Mott insulating phase. On the contrary,
nonvanishing $\langle X_i\rangle$ leads to charge fluctuations and describes non-Mott phases, which can still be insulating depending on whether the fermionic spectrum is gapped.
	
\section{\label{sec:results}Results}	

\subsection{Interacting phase diagram}
\begin{figure*}[t]
\includegraphics[width=0.9\textwidth]{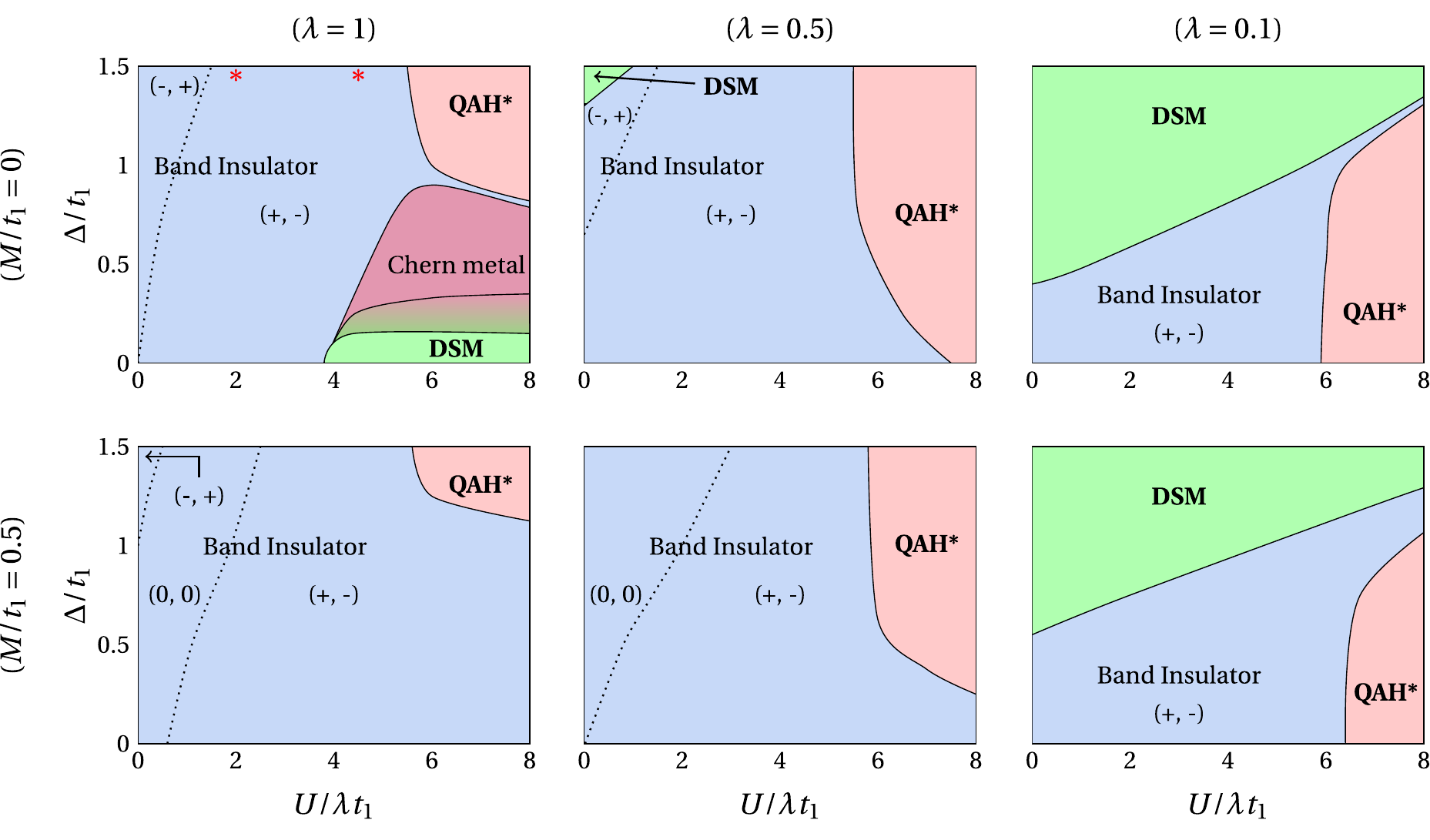}
\caption{Phase diagrams of the Haldane bilayer model obtained using slave rotor mean field theory as a function of the interaction strength
$U/\lambda t_1$ and the bias potential $\Delta/t_1$. Here, $\lambda < 1$ is the bandwidth
scaling factor of the correlated layer and $M$ is the degree of inversion symmetry breaking. There are five phases in total: `Band Insulator' phases which we distinguish by indicated Chern numbers for the valence bands, 
a `Dirac Semimetal' (DSM), 
a `Chern metal' with electron and hole pockets,
an intermediate metallic phase between the preceding two with the coexistence of Dirac cones and pockets (shown as the color gradient between the Chern metal and the DSM), and 
a fractionalized quantum anomalous Hall phase (QAH$^*$) in which the correlated layer undergoes a layer-selective Mott transition. The two stars in the upper left panel mark the points in the parameter space where we compute the electron spectral functions in section (\ref{sec:results}).}
\label{fig:pd}
\end{figure*}

The results of slave rotor theory are summarized in Fig.~(\ref{fig:pd}) where we plot phase diagrams of the bilayer Haldane model as we vary the interaction strength $U$ and the bias potential $\Delta$. The six panels in Fig.~(\ref{fig:pd}) correspond to different sets of $M$ and $\lambda$. We find the following phases: (1) band insulator, (2) Dirac semimetal, (3) Chern metal and (4) fractionalized quantum anomalous Hall insulator (QAH$^*$). Their properties are described below.
In our discussion of the band structure, note that each band is doubly degenerate in spin; below, we will describe one 
spin species unless otherwise mentioned explicitly.
	
	\subsubsection{Band insulator}
The noninteracting model is a trivial band insulator, and it continues to be a stable phase in a regime of the phase diagram at smaller $U$. In this phase, $\la X \ra \neq 0$, 
so the electrons in layer-2 are still well-defined excitations. The Chern numbers of the valence bands are shown in the parentheses; they sum up to zero so the insulator
is topologically trivial. The dashed lines separate three ground states with distinct Chern numbers $(-1, +1)$, $(+1, -1)$ and $(0, 0)$. The transition between them can be accomplished by gap closings between the valence bands to exchange their Chern numbers. This distinction is useful in understanding the effects the inversion breaking term and the evolution of the Chern bands across phase transitions.
	\subsubsection{Dirac semimetal (DSM)}
	\begin{figure}[h]
		\centering
    	\includegraphics[width=0.425\textwidth]{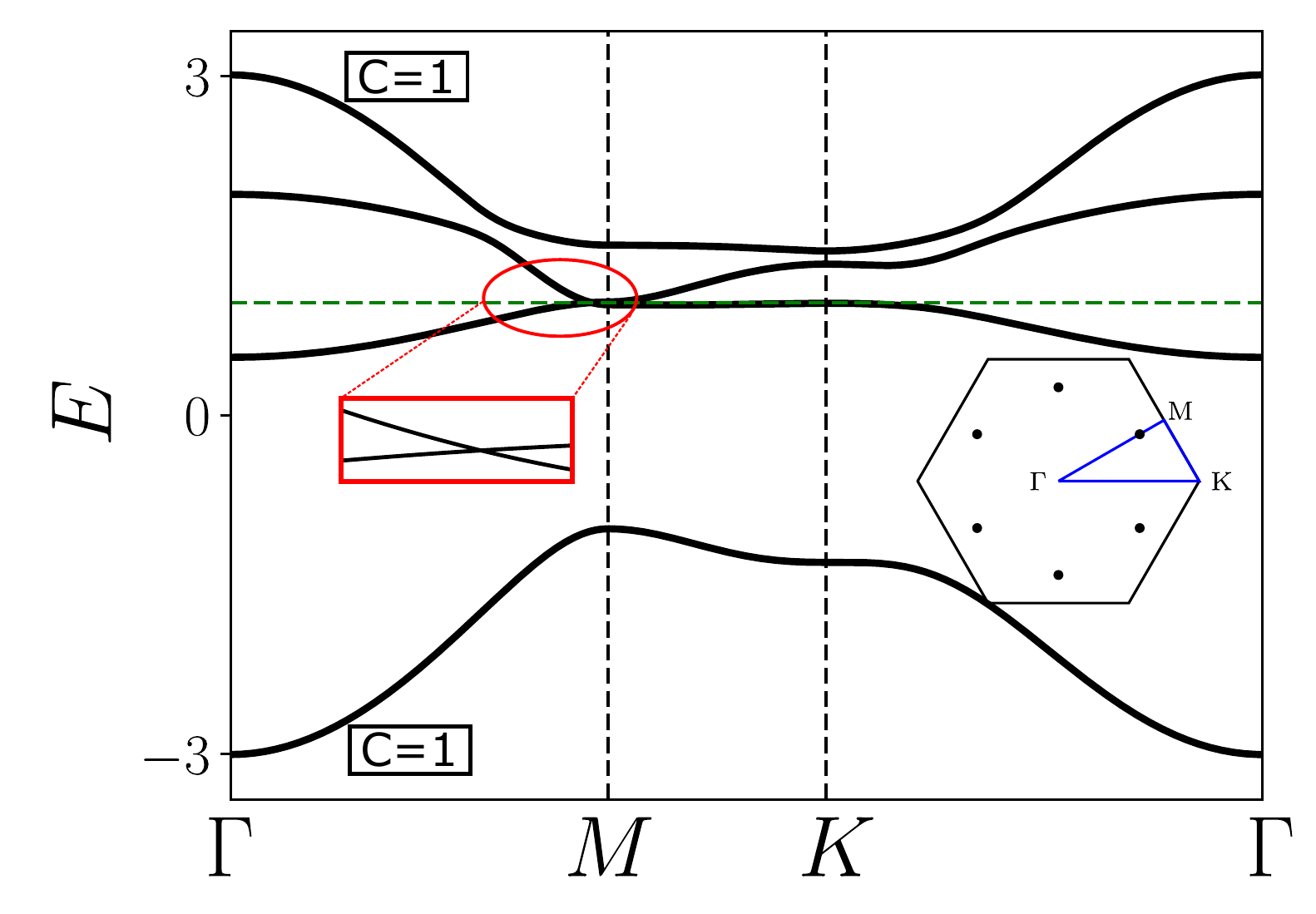}
		\caption{Fermionic band structure in the Dirac semimetal phase with six Dirac points lying on the $\Gamma$-M lines of the BZ as shown in the inset. The data is obtained at $M = 0$, $\lambda = 1$, $\Delta = 0$, $U = 6 t_1$ and $t_1 = 1$ in the upper left panel of Fig.~\ref{fig:pd}.}
		\label{fig:diracsm}
	\end{figure}
The DSM features 6 Dirac points in the BZ. Figure \ref{fig:diracsm} illustrates a band structure of the DSM. The Dirac points sit on the $\Gamma$-M lines in the BZ when $M = 0$. The inversion symmetry breaking term can move the Dirac points off the high symmetry lines. The transition from a band insulator to the DSM proceeds with the formation of gapless points at the BZ boundary (M points),  each of which then splits into a pair of Dirac cones moving towards the $\Gamma$ point. Similar to the Dirac cones in Section (\ref{sec:nonint}), the Dirac cones here are not protected by any symmetry and can be gapped out, e.g. by a third-nearest neighbor hopping. \\
	\subsubsection{Chern metal}
	\begin{figure}[t]
		\centering
    	\includegraphics[width=0.425\textwidth]{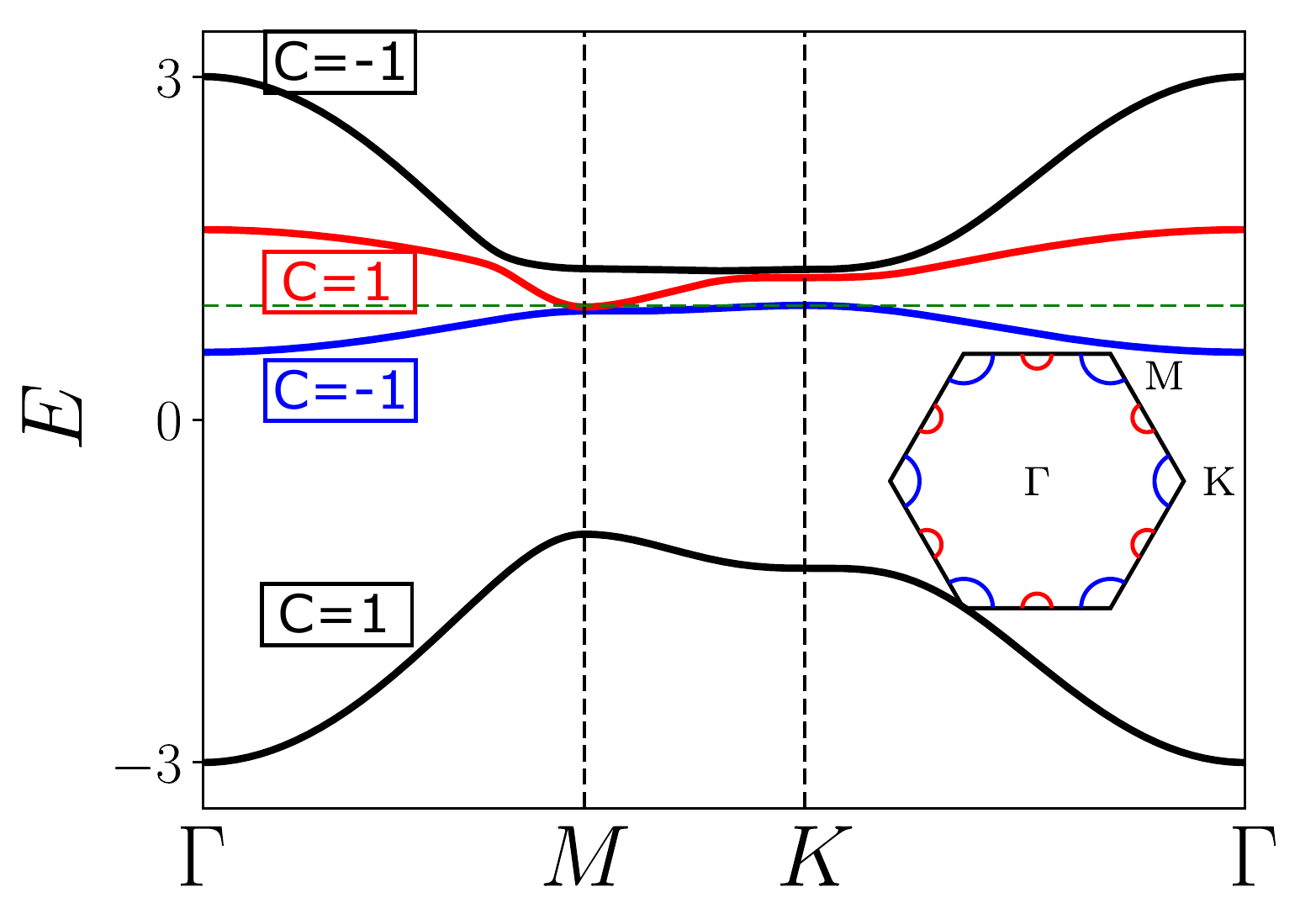}
		\caption{Fermionic band structure in the Chern metal phase. The Fermi level passes through a valence band and a conduction band, giving rise to hole pockets and electron pockets (see the inset: red for electron pockets and blue for the hole pockets.) The data is obtained at $M = 0, \lambda = 1, \Delta = 0.5t_1$, $U = 7t_1$ and $t_1 = 1$.}
		\label{fig:metallic-band}
	\end{figure}
In this phase, the band structure acquires electron pockets around M points and hole pockets around K and K' points as shown in Fig.~\ref{fig:metallic-band}, so it is
a compensated metal. Each band carries a nontrivial Chern number, which can lead to a finite Hall conductivity. A band structure calculation of the fermionic Hamiltonian on a cylinder with zigzag edges reveals no edge mode at the Fermi level, yet there are edge modes far below the Fermi level which start from the lower valence band and merge into the upper valence bands.

	The transition into the Chern metal can proceed in two ways from either a band insulator or a DSM. Starting from the band insulator, the band structure acquires electron and hole pockets and becomes a Chern metal. On the other hand, the transition from DSM passes through an intermediate metallic phase with electron and hole 
	pockets coexisting with Dirac cones. This phase is denoted by a color gradient in the upper left panel in Fig.~\ref{fig:pd}. The Chern metal phase appears after the 
	Dirac cones merge and gap out at the BZ boundary.
	
	\subsubsection{Fractionalized quantum anomalous Hall insulator}
This phase corresponds to a Mott phase in layer-2 in which the spin and charge of the correlated electrons dissociate. This kills the interlayer hybridization, 
resulting in an effective decoupling between the two layers.
Layer-1 is characterized by a nontrivial band topology with a bulk gap, electron-like excitations, and chiral electronic edge modes. The total Chern number is $+2$ (counting both
spins) which results in the quantization of electrical and thermal Hall conductivities \cite{ryu}.
	Layer-2 is described by a Mott phase with a topologically nontrivial spinon band structure. This corresponds to a topologically ordered chiral spin liquid as studied in Ref.~\onlinecite{slave_rotor_haldane_hubbard}. The chiral spin liquid has a gapped bulk spectrum, semion quasiparticles, and a chiral neutral gapless edge mode \cite{CSL1, CSL2}. To understand the neutral edge mode, one needs to go beyond the mean field treatment. The mean field spinon Hamiltonian suggests two spinon edge modes, but the ground state of the slave rotor can still have a finite overlap with unphysical states (those  which violate the constraint (\ref{eq:hilbert-constraint})). Thus, one needs to consider a projection onto the physical Hilbert space. As argued in Ref.~\onlinecite{chiral_central_charge}, the two spinon modes can be identified with a gapless charged mode and a gapless neutral mode, the former of which will be gapped out upon the projection, leaving only one gapless neutral mode at the boundary. Its neutrality leads to a zero contribution to the electrical Hall effect while contributing a quantized thermal Hall conductivity of one unit quantum \cite{ryu}.
	
	The properties of the total system can be summarized below. In the bilayer, the quasiparticle exciations consist of electrons and semions. At the boundary, there are gapless chiral charged modes and a counterpropagating neutral mode. The total electric Hall conductivity is $\sigma_{xy} = 2\frac{e^2}{h}$ which solely arises from layer-1, while the thermal Hall conductivity $\kappa_H$ equals to $+1$ quantized unit of the Hall conductivity ($+2$ and $-1$ from layer-1 and layer-2 respectively). The relation between $\kappa_H$ and $\sigma_{xy}$ violates the Wiedemann-Franz law without having a vanishing or a fractional electric Hall conductivity (like those in spin liquids and fractional quantum Hall liquids \cite{thermal_hall_fqh}). QAH$^*$ is similar to fractionalized Fermi liquids (FL$^*$), proposed in Ref.~\onlinecite{frac_FL}, in which the spins and the electrons of a Kondo lattice model are effectively decoupled, where spins are fractionalized and form a spin liquid phase, while the electrons form a Fermi liquid.

We have found that the QAH$^*$ phase occupies  a significant portion of the phase diagram. 
In the topological bootstrap limit $\lambda \ll 1$, the QAH$^*$ phase can arise straightforwardly -- without the bias potential -- upon increasing the interaction strength. When one increases the strength of $M$, the phase diagram changes quantitatively in that the QAH$^*$ phase is pushed to the right as one requires a larger $U$ to drive the system into a Mott phase on account of an increased tendency to a charge imbalance between the two sublattices. As one departs from the topological bootstrap limit, a Mott phase requires a bias potential $\Delta > 0$ to compensate for the energy cost by the Hubbard interactions in order to hold electrons in layer-2 to a half-filled density.

\begin{figure}[t]
		\centering
    	\includegraphics[width=0.425\textwidth]{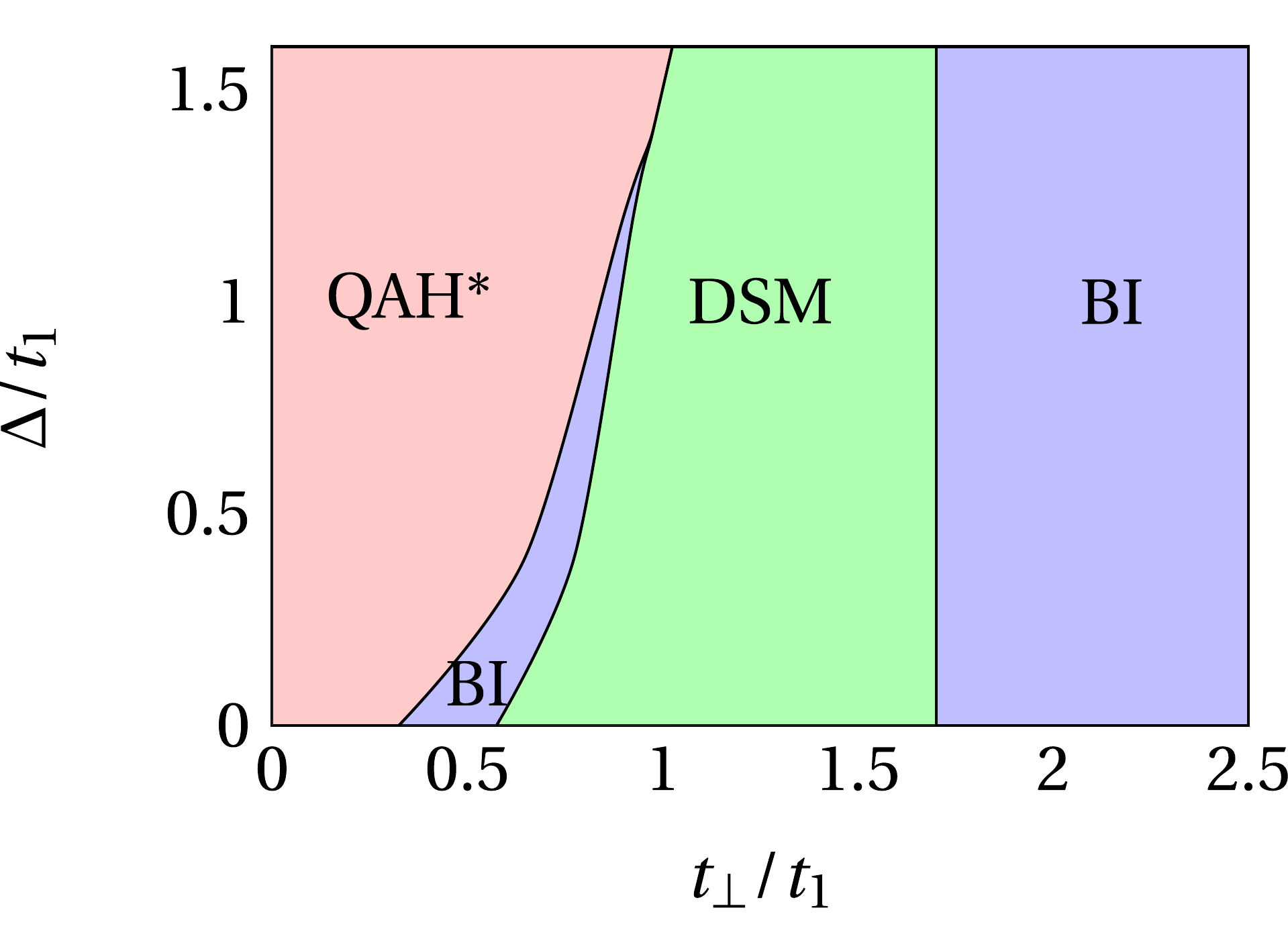}
		\caption{A phase diagram depicting the effects of hybridization $t_{\perp}$ on QAH$^*$. The system becomes a trivial band insulator (BI) at large hybridizations, which is identifiable with a Kondo insulator in the strong-coupling limit. The two BI regions are physically the same in terms of their Chern numbers and the presence of a gap in the band structure. The intermediate Dirac semimetal phase may be gapped out upon adding other microscopic terms. Thus it is possible to have a direct transition from QAH$^*$ into BI. (The phase diagram is obtained at $U = 8.0 \lambda t_1, \lambda = 0.5, M = 0$.)}
		\label{fig:tperp}
	\end{figure}
Upon increasing the strength of the hybridization $t_{\perp}$, QAH$^*$ passes through intermediate phases before eventually becoming a topologically trivial band insulator (BI). BI consists of very flat fermionic bands hybridizing with dispersive bands, which can be identified with a Kondo insulator where every spin binds an electron. Phase diagram in Fig. \ref{fig:tperp} illustrates the hybridization effects on QAH$^*$. The intermediate phase being the DSM can be gapped out by adding other microscopic terms to the Hamiltonian. Thus it is possible for QAH$^*$ to have a direct transition into the BI phase.

\subsection{\label{sec:spec-func}Edge electron spectral function}
\begin{figure}[t]
		\centering
    	\includegraphics[width=0.5\textwidth]{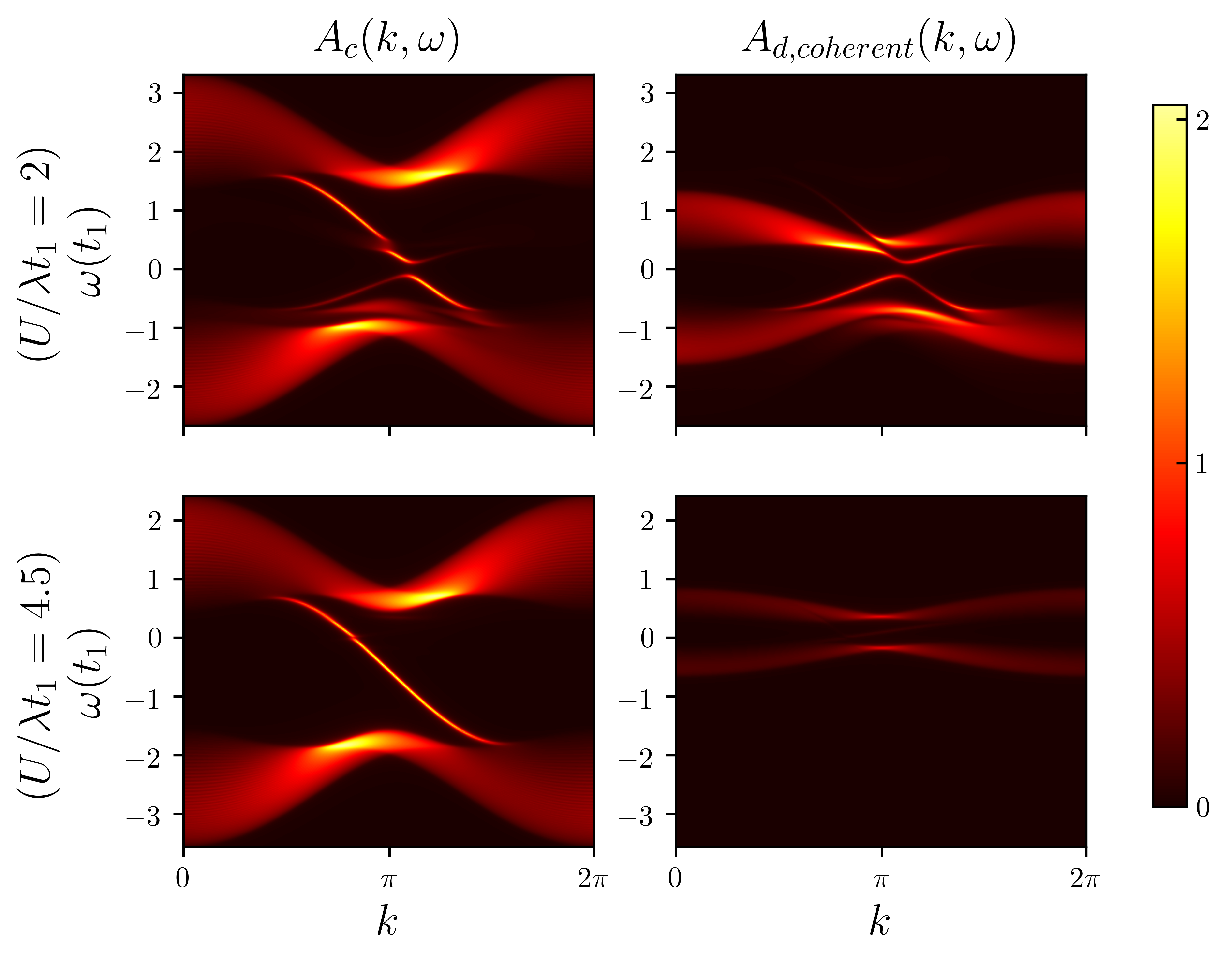}
		\caption{Spectral functions of electrons in layer-1 and layer-2 (left and right columns) near a zigzag boundary of a cylinder at $U  = 2t_1$ (deep in the band insulator phase) and $U = 4.5 t_1$ (close to the transition to QAH$^*$), marked by stars in the upper left panel of Fig.~\ref{fig:pd} with $\Delta = 1.5t_1$. Away from the transition, the hybridization is still profound, so the edge modes from layer-1 and layer-2 are hybridized, leading to a gap at $\omega = 0$. One can still identify the remnant of a left moving edge mode in the upper left panel and a right moving edge mode in the upper right panel. On the contrary, as the system approaches the QAH$^*$, the coherent part of the spectral function of the correlated electron fades away since the hybridization is increasingly suppressed, leaving $A_c$ almost unaffected.}
		\label{fig:spec-func}
	\end{figure}
The ground state wavefunction in slave rotor mean field theory is a direct product of spinon and rotor wavefunctions,
which allows us to determine the electron Green function $G_d(x, \tau) \equiv \langle \text{T}_{\tau}d_x(\tau)d^{\dagger}_0(0)\rangle$. 
In real space, $G_d$ is the product of spinon and rotor Green functions, $G_d(x, \tau) = G_f(x, \tau) \times G_X(x, \tau)$, so in momentum space it becomes a convolution:
\begin{align}
G_d(\bk, i\omega_n) &= \sum_{\bq, i\Omega_n} G_f(\bq, i\Omega_n) G_X(\bk-\bq, i\omega_n - i\Omega_n).
\end{align}
Here, $G_X(\bp, i\nu_n)=Z\delta_{\bp, 0}\delta_{\nu_n, 0} + \tilde{G}_X(\bp, i\nu_n)$, where the first part is the contribution from zero momentum and frequency $Z \sim |\langle X_i\rangle|^2$. Then $G_d(\bk, i\omega_n) = ZG_f(\bk, i\omega_n) + \tilde{G}_d(\bk, i\omega_n)$, where the two terms are the coherent and incoherent parts respectively. The coherent part provides a sharp contribution to the spectral function $A_d \propto \text{Im}G_d(\bk, i\omega_n \rightarrow \omega + i\epsilon)$, while the incoherent part is smeared out by the convolution.

	As suggested by the phase diagrams, the QAH$^*$ phase is only unstable to the band insulator, so it is interesting to see how the spectral functions change as the system approaches the QAH$^*$. We compute the spectral functions at two points marked by the two stars in the upper left panel of Fig.~\ref{fig:pd}; one 
	point is deep in the band insulator, while the other is close to a transition to QAH$^*$. Figure \ref{fig:spec-func} shows the spectral functions of the electrons in layer-1 (layer-2) denoted by $A_{c(d)}$, while the subscript ``coherent" denotes the coherent part. They are obtained from a calculation on a cylinder with zigzag edges using the self-consistent bulk Hamiltonians. In the calculation, the Green functions are computed from electron operators located close to a boundary of the system.

	Deep in the band insulator phase, the gap at $\omega = 0$ in the edge states is the result of a hybridization between the electronic and the spinon edge modes (upper panels of Fig.~\ref{fig:spec-func}). 
	The signature of the left moving and right moving edge modes are still fairly apparent despite the hybridization effect. In contrast, near the transition to 
	QAH$^*$ (lower panels), the hybridization is strongly suppressed, so the edge mode in the bottom left panel is almost unaffected, and thus resembles that of a 
	Chern insulator. 
	Meanwhile, $A_{d, \rm coherence}$ progressively fades away in both the in-depth states (the continuum) and the edge state (in-gap state).
	We conclude the section by noting that such a small hybridization near the phase boundary may lead to an incorrect characterization of the system in experiments. For instance, if electrons can undergo Landau-Zener tunnelling across the hybridization gap, one might incorrectly conclude that the system exhibits a QAH effect 
despite the trivial band topology.\\

\section{Summary}
We have studied a bilayer Haldane model under the effect of electron correlations as a lattice Anderson version of the topological bootstrap. Using slave rotor theory, we have explored the phase diagram of this model, and have found a fractionalized quantum anomalous Hall insulator arising from a trivial insulator which undergoes a layer-selective Mott transition in the strongly correlated regime. This phase has coexisting electronic and semionic bulk excitations and is predicted to exhibit a combination of a quantized electrical Hall effect and a quantized thermal Hall effect which violates the Wiedemann-Franz law due to fractionalization. The hybridization can drive QAH$^*$ into a topologically trivial insulator, which can be viewed as a Kondo insulator in the strong-coupling limit. However, it is not a direct transition in the current model; a direct transition between QAH$^*$ and the trivial insulator may be achieved by adding other microscopic terms, which we leave for future studies.

\acknowledgements

This research was supported by NSERC of Canada. The author would like to thank Arun Paramekanti and Timothy Hsieh for very helpful discussions. He also would like to thank Joseph Maciejko for useful comments.

\bibliography{references} 
\onecolumngrid
\clearpage
\appendix

\section{Origin of the Dirac points}
\label{appdx: origin-dirac-points}

The Dirac points arise at the momenta where the fermionic Bloch Hamiltonian of $H_{f,c}$ can be written in a special form: \emph{The first diagonal block differs from the second one by a rescaling and a constant, } namely:
\begin{align}
	\label{eq:fermionic-bloch-hamiltonian}
	  \begin{split}
		&\mathcal{H}_{c, f}(\bk) ={} \begin{pmatrix}
\alpha + \gamma \mathcal{H}_1(\bk) & V \\
V & \mathcal{H}_1(\bk)
\end{pmatrix}\\
		&={} V\tau_x + \frac{\alpha + (\gamma - 1)\mathcal{H}_1(\bk)}{2}\tau_z + \frac{\alpha + (\gamma + 1) \mathcal{H}_1(\bk)}{2} \tau_0,
	  \end{split}
	\end{align}
where $\mathcal{H}_1$ is a two-by-two matrix, $V, \alpha, \gamma$ are constant and $\tau$'s are Pauli matrices acting on the layer index. To see this, we first start with the general form of the Bloch Hamiltonian with nearest and next nearest neighbor hoppings in the basis of $(f_{\bk A\sigma}, f_{\bk B\sigma}, c_{\bk A\sigma}, c_{\bk B\sigma})^T$,
\begin{align}
	  \begin{split}
		\mathcal{H}_{c, f}(\bk) &={} \begin{pmatrix}
\alpha + \beta \mathcal{H}_1(\bk) + \gamma \mathcal{H}_2(\bk) & V \\
V & \mathcal{H}_1(\bk) + \gamma' \mathcal{H}_2(\bk)
\end{pmatrix},
	  \end{split}
	\end{align}
where $\mathcal{H}_1$ and $\mathcal{H}_2$ are $2\times 2$ matrices originating from the nearest and next nearest neighbor hoppings respectively. $\alpha, \beta, \gamma, \gamma'$ are constant, while $V$ is the hybridization term. We have switched the order of the layer for simplicity, and the $f^{\dagger}$ operator can either be electron or spinon in layer-2.

	\begin{figure}[t]
		\centering
    	\includegraphics[height=0.25\textwidth]{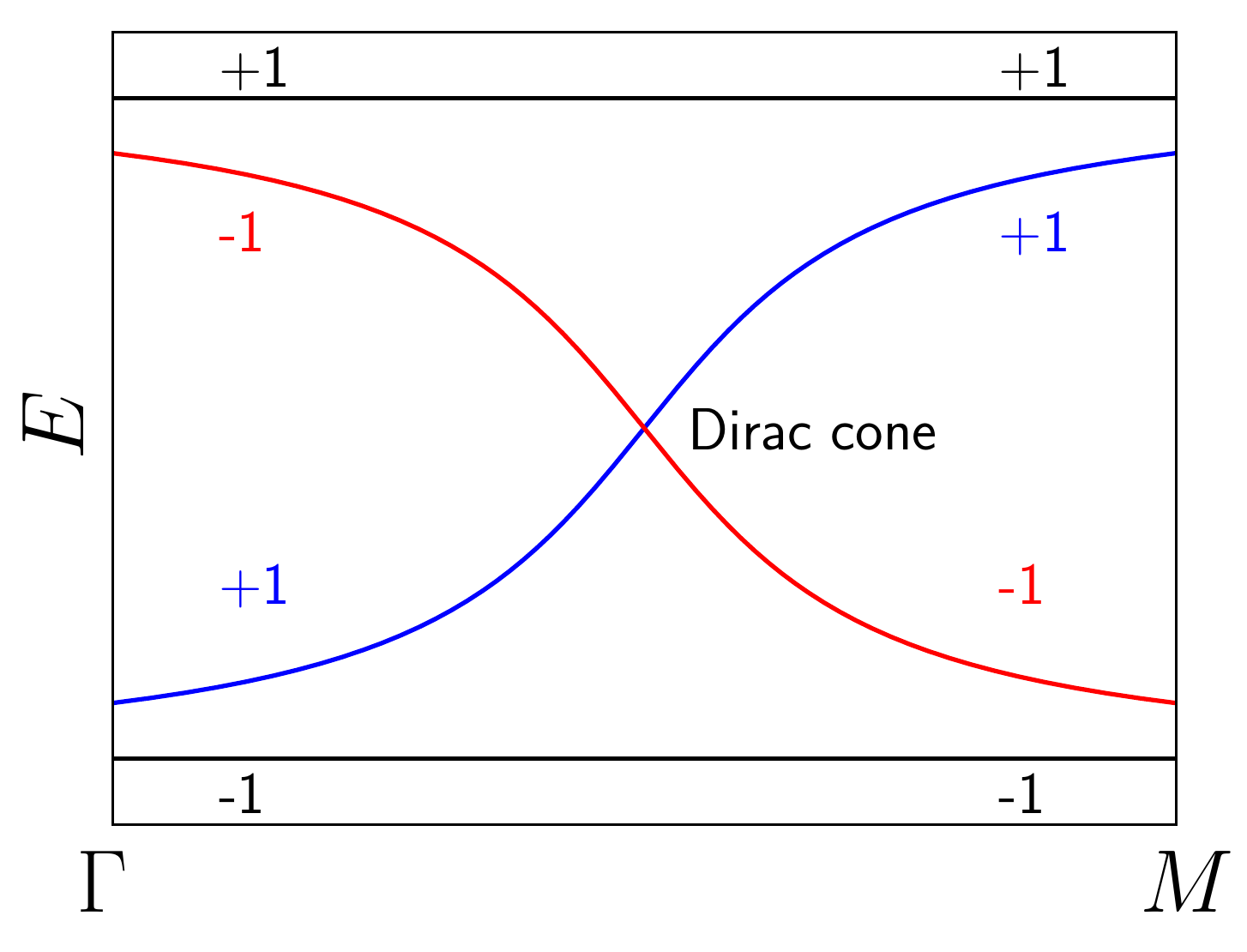}
		\caption{Schematic band structure along a $\Gamma$-M line illustrating the presence of a Dirac cone. The $\pm 1$ are the eigenvalues of the matrix $\mathcal{U(\bk)}$}
		\label{fig:origin-dirac-cone}
	\end{figure}
	
The special form in Eq.(\ref{eq:fermionic-bloch-hamiltonian}) is obtained when $\mathcal{H}_2(\bk) = 0$. In the Haldane model, the block $\mathcal{H}_2(\bk) = -t_2\sigma_z \sum_{\mu} \sin(\bk \cdot \mathbf{b}_{\mu})$ is diagonal and real-valued, where $\sigma$ acts on the sublattice index. $\mathbf{b}_{\mu}$'s are the three vectors connecting a site with its next nearest neighbors in the honeycomb lattice. Hence the special form can occur on a (set of) curve, i.e. a 1D object, in the BZ as the result of solving an equation with two unknowns. When the inversion breaking term is absent, the 1D object consists of the $\Gamma$-M lines. We should restrict the discussion below to this special case for simplicity. Consider a matrix,
\begin{align}
		\mathcal{S}(\bk) &= V\tau_x + \frac{\alpha + (\gamma - 1)\mathcal{H}_1(\bk)}{2}\tau_z,
	\end{align}
which is just the first two terms in (\ref{eq:fermionic-bloch-hamiltonian}). Clearly $\mathcal{S}(\bk)$ commutes with $\mathcal{H}_{c, f}(\bk)$ so the Bloch wavefunctions on the $\Gamma$-M lines are also eigenfunctions of $\mathcal{S}(\bk)$. Define another matrix which take the signs of the eigenvalues of $\mathcal{S}(\bk)$, $\mathcal{U}(\bk) = \text{Sgn}(\mathcal{S}(\bk))$. Then the formation of the Dirac cones can be understood as the crossing of the bands with $\pm 1$ eigenvalues of $\mathcal{U}(\bk)$ along the $\Gamma$-M lines as illustrated in Fig.~\ref{fig:origin-dirac-cone}. Another way to understand the Dirac cones is to impose another constraint on the eigenvalues of $\mathcal{H}_{c, f}(\bk)$ of the special form such that the middle two bands have an equal energy. We have two variables, $k_x$ and $k_y$, to tune in order to satisfy the two equations, hence the Dirac cones can exist at multiple points on the $\Gamma$-M lines of the BZ.
	
	The form of $\mathcal{S}(\bk)$ can hint a way to gap out the Dirac points. One example is to introduce third nearest neighbor hoppings, forbidding the special form thereby, gapping out the Dirac cones.\\

\section{Nonlinear sigma model representation of rotor mean field Hamiltonian}
\label{appdx: sigma-model}
Here we briefly outline the computation of the rotor bond mean fields in (\ref{eq:uniform-ansatze-nn})-(\ref{eq:uniform-ansatze-hyb}). They are computed using a Euclidean action constructed from the rotor Hamiltonian $H_{\theta}$ which is then represented by a nonlinear sigma model. The nonlinear sigma model representation is quadratic and is used to compute the bond mean fields.
The Euclidean action constructed from $H_{\theta}$ is given by
	\begin{align}
	  \begin{split}
	    S_E[L, \theta] =&{} \int_0^{\beta} d\tau \left[ \sum_i \left(-iL_i\partial_{\tau}\theta_i + \lambda_i L_i + \frac{U}{2}L_i^2\right) + \left(\sum_{\langle ij\rangle} -2t_1 F_{nn} e^{-i\theta_i}e^{i\theta_j} - \sum_{i} 2t_{\perp}F_{\perp, i} e^{-i\theta_i} + \text{c.c.}\right) \right]\\
		&{}+ \int_0^{\beta}d\tau \left(\sum_{\langle\langle ij\rangle\rangle} - 2t_2F_{nnn, i}e^{-i\theta_i}e^{i\theta_j} e^{-i\nu_{ij}(\phi + \varphi_i)} + \text{c.c.} \right),
	  \end{split}
	\end{align}
	where the partition function $\mathcal{Z} = \int \mathcal{D}\theta\mathcal{D}L \exp(-S_E)$. Integrating out the angular momentum field, we obtain
	\begin{align}
	  \begin{split}
	    S_E[L, \theta] =&{} \int_0^{\beta} d\tau \left[\sum_i \left(\frac{(\partial_{\tau}\theta_i)^2}{2U} + \lambda_i \frac{i\partial_{\tau}\theta}{U}\right) + \left(\sum_{\langle ij\rangle} -2t_1 F_{nn} e^{-i\theta_i}e^{i\theta_j} - \sum_{i} 2t_{\perp}F_{\perp, i} e^{-i\theta_i} + \text{c.c.}\right)\right]\\
		&{} + \int_0^{\beta}d\tau \left[\sum_{\langle\langle ij\rangle\rangle} - 2t_2F_{nnn, i}e^{-i\theta_i}e^{i\theta_j} e^{-i\nu_{ij}(\phi + \varphi_i)} + \text{c.c.} \right].
	  \end{split}
	\end{align}
	Replacing the phase factor by a sigma field $X = e^{i\theta}$ whose constraint $|X_i|^2 = 1$ is imposed using two Lagrange multipliers $\rho_{A(B)}$. One arrives at a nonlinear sigma model of the rotor Hamiltonian.
	\begin{align}
	  \begin{split}
		S_E[X^*, X] =&{} \int_0^{\beta} d\tau \left[\sum_i \left(\frac{|\partial_{\tau}X_i|^2}{2U} + \frac{\lambda_i}{2U} (X^*_i\partial_{\tau}X_i - \partial_{\tau}X^*_iX_i)\right) +\left(\sum_{\langle ij\rangle} -2t_1 F_{nn} X^*_iX_j - \sum_{i} 2t_{\perp}F_{\perp, i} X^*_i + \text{c.c.}\right)\right]\\
		&{} + \int_0^{\beta}d\tau \left[ \sum_{\langle\langle ij\rangle\rangle} \left(- 2t_2F_{nnn, i}X^*_iX_j e^{-i\nu_{ij}(\phi + \varphi_i)} + \text{c.c.} \right) + \sum_i  \rho_i X^*_iX_i \right].
	  \end{split}	
	\end{align}
	This is a quadratic action which can be used to compute the bond mean fields. In the Fourier space $(\bk, i\omega_n)$, the action is given by
	\bea
	S_E &=& \sum_{\bk, n, s, s'} X^*_{\bk, n, s}\mathcal{S}_{s, s'}(\bk, \omn) X_{\bk, n, s'} - \sum_{s} 2t_{\perp}F_{\perp, s}\sqrt{\beta N_c}(X_{0, 0, s} + \rm c.c.)\\
	\mathcal{S}_{s, s'}(\bk, \omn) &=&	\begin{pmatrix}
	\frac{\omn^2}{2U} + \frac{\lambda_A}{U}i\omn + \rho_A& \\
	 & \frac{\omn^2}{2U} + \frac{\lambda_B}{U}i\omn + \rho_B
	\end{pmatrix} + \mathcal{H}^X_{s, s'}(\bk),
	\eea
	where $N_c$ is the number of unit cell in layer-2; $s, s'$ are sublattice indices and $\mathcal{H}^X_{s, s'}(\bk)$ originates from the inplane hopping terms in Fourier space which does not depend on the frequency. We have added the inversion symmetry breaking term (only present explicity in the fermionic sector), which leads to two sigma fields to account for the charge imbalance in a generic case. Two $\lambda$'s and two $\rho$'s
are needed as a consequence.

	The following propagator is essential for solving the self-consistent conditions:
	\bea
	\label{eq:propagator}
		\langle X_{\bk, \omn, s}X^{\dagger}_{\bk, \omn, s'} \rangle &=& \mathcal{S}^{-1}_{s, s'}(\bk, \omn) + \delta_{\bk, 0}\delta_{\omn, 0}\langle X_{0, 0, s} \rangle \langle X^*_{0, 0, s'} \rangle,
	\eea
	where $\langle X_{0, 0, s} \rangle = \sum_{s'}2 t_{\perp} \sqrt{\beta N_c}\ \mathcal{S}^{-1}(0, 0)_{s, s'}F_{\perp, s'}$. To satisfy the sigma field constraints, we impose the following conditions
	\bea
		1 &=& \frac{1}{N_c} \sum_{i} \langle X^{\dagger}_{is}(\tau = \epsilon)X_{is}(0) \rangle \nonumber\\
		&=& \frac{1}{\beta N_c} \sum_{\bk, n} \langle X^{\dagger}_{\bk\omn s} X_{\bk\omn s}\rangle e^{-i\omn \epsilon},
	\eea
	where $\epsilon \rightarrow 0^+$ is to keep the correct ordering of the operators. To compute the frequency sum, we use the residue theorem to convert the sum over the poles of the Bose function $n_B(z) = \frac{1}{e^{\beta z} - 1}$ into a pole sum of a function $f(\bk, z)  \overset{i\omn \rightarrow z}{=}\langle X^{\dagger}_{\bk\omn s}X_{\bk \omn s}\rangle$. We compute the poles of $f(\bk, z)$ numerically and obtain the frequency sum. The Hilbert space constraints can be expressed similarly in terms of the propagator in Eq.~(\ref{eq:propagator}). They involve the expectation values of the angular momentum obtained from a Heisenberg equation of the rotor operator $\partial_{\tau} X_{is} = \left[H_{\theta}, X_{is}(\tau)\right]$ and the commutator $\left[L, e^{i\theta}\right] = e^{i\theta}$.
	\bea
		L_{is} &=& -\frac{\lambda_s}{U} - \frac{1}{2} + \frac{1}{U} X^{\dagger}_{is}\frac{\partial X_{is}}{\partial \tau}.
	\eea
	Likewise, the bond mean fields in (\ref{eq:uniform-ansatze-nn})-(\ref{eq:uniform-ansatze-hyb}) can also be computed using the pole summation procedure described above.
	
\end{document}